\documentclass[conference]{IEEEtran}
\IEEEoverridecommandlockouts
\usepackage{cite}
\usepackage{amsmath,amssymb,amsfonts}
\usepackage{algorithmic}
\usepackage{graphicx}
\usepackage{textcomp}
\usepackage{xcolor}

\usepackage{caption}
\usepackage{subcaption}
\usepackage[colorlinks,bookmarksopen,bookmarksnumbered,citecolor=blue,urlcolor=blue]{hyperref}
\def\BibTeX{{\rm B\kern-.05em{\sc i\kern-.025em b}\kern-.08em
    T\kern-.1667em\lower.7ex\hbox{E}\kern-.125emX}}
\begin{document}

\title{A Chronological Analysis of the Evolution of SmartNICs\\}

\author{\IEEEauthorblockN{1\textsuperscript{st} Olasupo Ajayi}
\IEEEauthorblockA{\textit{CAESAR Laboratory,}\\
\textit{Electrical \& Computer Engineering Department,} \\
\textit{Queen's University, Kingston,}\\
ON, K7L3N6, Canada \\
0000-0001-6583-3749
}
\and
\IEEEauthorblockN{2\textsuperscript{nd}Ryan Grant}
\IEEEauthorblockA{\textit{CAESAR Laboratory,}\\
\textit{Electrical \& Computer Engineering Department,} \\
\textit{Queen's University, Kingston,}\\
ON, K7L3N6, Canada \\ 
0000-0002-0163-3892
}
}

\maketitle

\begin{abstract}
Network Interface Cards (NICs) are one of the key enablers of the modern Internet. They serve as gateways for connecting computing devices to networks for the exchange of data with other devices. Recently, the pervasive nature of Internet-enabled devices coupled with the growing demands for faster network access have necessitated the enhancement of NICs to Smart NICs (SNICs), capable of processing enormous volumes of data at near real-time speed. However, despite their popularity, the exact use and applicability of SNICs remains an ongoing debate. These debates are exacerbated by the incorporation of accelerators into SNIC, allowing them to relieve their host's CPUs of various tasks. In this work, we carry out a chronological analysis of SNICs, using 370 articles published in the past 15 years, from 2010 to 2024, to gain some insight into SNICs; and shed some light on their evolution, manufacturers, use cases, and application domains.
\end{abstract}


\section{Introduction}
Smart Network Interface Cards (SNICs) are a type of network interface card (NIC) that are equipped with a system-on-chip (SoC) capable of running full-fledged operating systems and can be programmed to handle various tasks, thus offloading such tasks from the host's Central Processing Unit (CPU) \cite{Miano, nvdev}. 

Recently, several terminologies have emerged around the concept of SNICs, including Network Accelerators, Data Processing Unit (DPU), and FPGA-NIC (Field Programmable Gateway Array - NIC). There has also been a growing debate about the exact purpose and use cases of these devices. In many contexts, especially in High Performance Computing (HPC), SNICs are considered accelerators that are used to offload various tasks off the main CPU. The concepts of network and workload offloading are well discussed in literature hence, not repeated here. Rather, in this work, we carry out a chronological analysis of the concept of SNICs, exploring scholarly publications spanning a 15-year period from 2010 to 2024. The purpose and main contributions of this work are as follows:

\begin{itemize}
    \item To gain insights into how SNICs are being used in different scenarios, by exploring applications and use cases reported in literature. 
    \item Identify the main research areas where SNICs have been and are being applied. 
    \item Curate a dataset of published works on SNICs spanning a 15-year period.
    \item Identify the main manufacturers of SNICs and how their products have evolved over the past decade and a half. 
\end{itemize}

\section{Background}
Although the  Transmission Control Protocol/Internet Protocol (TCP/IP) and Open Systems Interconnection (OSI) stacks have received a lot of attention, NICs are undoubtedly the unsung heroes of modern networks. A NIC (or network adapter) is a hardware component that connects a computing device to a network and allows the device to receive or send data to other devices on the network. NICs operate in both the first and second layers (physical and data link) of the OSI stack or layer 1 (link layer) of the TCP/IP protocol stack, implementing the IEEE 802.3 (wired Ethernet) and IEEE 802.11 (wireless) communication standards.   

\begin{figure}%
\begin{subfigure}{0.5\textwidth}
         \includegraphics[width=\textwidth]{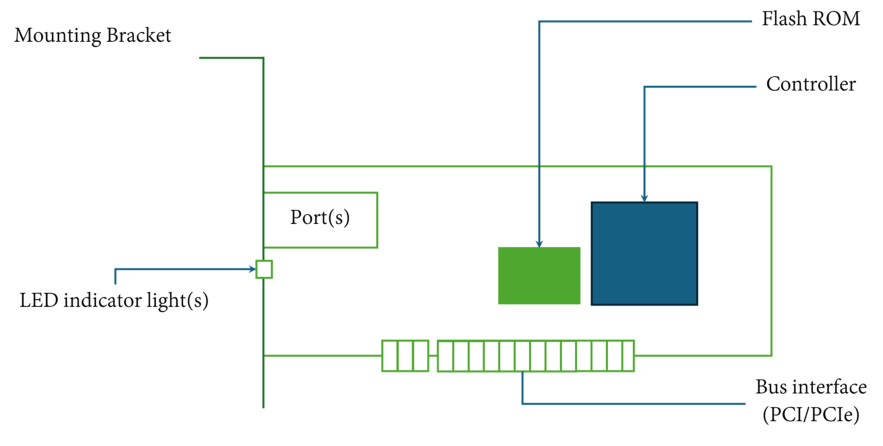}
         \subcaption{Generic Network Interface Card Layout}
         \label{fig1}
     \end{subfigure}
    \begin{subfigure}{0.5\textwidth}
         \includegraphics[width=\textwidth]{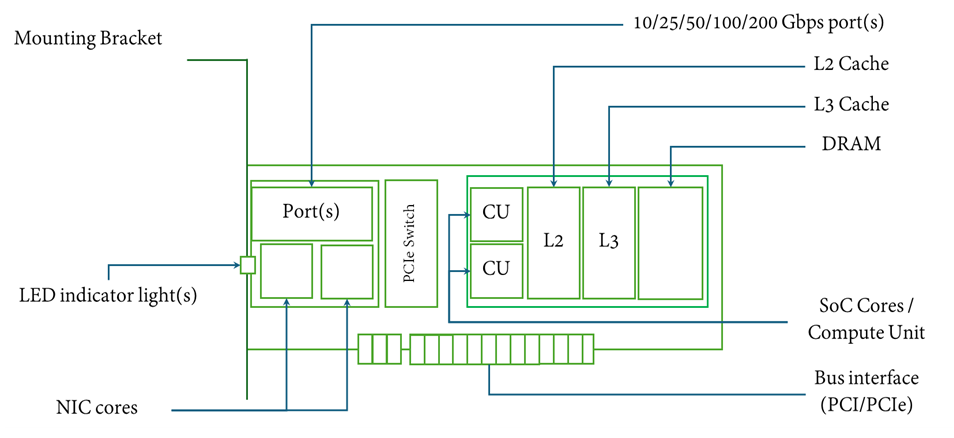}
         \subcaption{Generic SNIC}
         \label{fig2}
     \end{subfigure}
     \caption{Network Interface Cards}
     \label{fig1-2}
\end{figure}

NICs are responsible for host addressing within a network (via Media Access Control or MAC address), data transmission (between its host device and the network), transmission control (including congestion avoidance and prevention of packet collisions), and error detection and correction (via checksums and redundancy checks). Fig. \ref{fig1-2} shows a graphical comparison of the basic layouts of a classic NIC and a smart NIC (SNIC).

Computer networks have been growing at an exceptional rate in recent years. This is mainly due to the increase in global internet usage, which, as at the time of writing, stands at approximately 66\% of the world’s population \cite{drp}, and the ubiquity of smart phones and connected devices. These factors have led to the evolution of NICs from mere network connection points to more advanced devices fitted with specialized circuitry capable of handling higher data transmission rates, faster processing, and more functionalities, called Smart NICs (SNICs).  

From a functionality standpoint, SNICs can be classified into three categories - Application Specific Integrated Circuit (ASIC) based, Field Programmable Gateway Array (FPGA) based, and those that incorporate SoCs, called Data Processing Units (DPUs). ASIC-based SNICs, as the name suggests, are designed for a specific purpose, such as offloading network tasks off the CPU. These NICs are cheaper but not flexible, hence, cannot easily adapt to network needs. FPGA-based SNICs, on the other hand, are flexible and can be programmed for multiple network tasks. DPUs (SoC-based SNICs) have built-in processing units (ARM processors), memory, and host interface, in addition to ASIC or FPGA processing elements. They can be considered synonymous to edge or single-board computers and are highly programmable using C/C++ and/or Programming Protocol-independent Packet Processors (P4) programming languages. DPUs often have multiple accelerators and can be programmed to offload several tasks from their host’s CPU \cite{Miano, nvdev, ychi}. 

\section{Related Works}
A comprehensive review of SNICs was carried out in \cite{chpt2}. The work defined “SmartNICs” as programmable NICs, then compared various SNICs based on their design, hardware architecture, and type (ASIC, FPGA, SoC, Hybrid). A survey of the programmability of SmartNICs, using P4, C/C++, and Nvidia's DOCA was also done by the authors. Finally, the work considered some common SmartNICs, specifically Nvidia’s Bluefield and Netronome’s NFP series, reviewed research publications where they were used, and discussed some of the open challenges and limitations of SmartNICs in general. 

In \cite{tristan}, a survey of SNICs was performed from both industrial and research perspectives. The authors reviewed several articles in an effort to propose a universal definition of the term “SmartNIC”. This was then followed by an exploration of the various architectures, i.e., ASIC-based, FPGA-based, and SoC-based SNICs. Finally, the work compared two products, Intel’s Programmable Acceleration Card (PAC) N3000 and Nvidia’s Bluefield-2, and their performance in mitigating Distributed Denial of Service (DDoS) attacks.

In a similar work, \cite{kfoury} surveyed SNICs holistically. The authors began by comparing classic NICs with SmartNICs and then showed the evolution from the former to the latter. The work grouped SNICs into five categories, namely, ASIC-based, FPGA-based, ASIC+FPGA hybrid, ASIC+SoC and FPGA+SoC. Using a taxonomy as a guide, the authors compared various SNICs based on their programmability, CPU architecture, and traffic switching abilities. Task offloading using SNICs was also discussed and grouped into security, network, storage, and computing tasks. The paper was concluded with a discussion on the open challenges of SNICs. 

An evolutionary trend analysis of SNICs was carried out in \cite{bodd}. SNICs from major manufacturers were reviewed, as well as their applications as workload accelerators in various domains, including telecommunications, storage, artificial intelligence, and edge processing. Similarly, \cite{will23} also explored the potential of using SNICs as general-purpose accelerators in HPC. They explored the offloading mechanisms of Nvidia’s Bluefield-2 DPU using Pennant and BigSort benchmarking applications. 

This work, unlike other related works, does not focus on the architectures, taxonomy, or comparison of offloading capabilities of SNICs in HPC; instead, we carry out a systematic chronological analysis of SNICs, using published articles, to provide insights into their evolution and real-world applications over a 15-year period. To the best of our knowledge, this work is the first of its kind with respect to SNICs.  

\section{Methodology}
Being a systematic review, this work followed the PRISMA guidelines \cite{prisma}, with the following steps taken to carry out the analysis.
\begin{itemize}
    \item \textbf{Step 1 – Data collection:} data on SNICs related publications were curated from IEEE Xplore repository. “DPU”, “SmartNIC”, and “FPGA-NIC” were used as keywords, with the search limited to articles published in the 15 years between 2010 and 2024. 

    \item \textbf{Step 2 – Filtration / screening:} Step 1 resulted in about 490 articles, which were then passed through two filtration phases. In the first filtration phase, we read the abstract and index or key words of each article to filter out irrelevant articles. 26 articles were excluded through this process. In the second phase, we read the entire content of each of the remaining articles. During this phase, we were able to exclude articles relating to smart switches and other solutions that were also termed data processing units (DPU) but referred to systems designed for other forms of end-to-end data processing. Finally, we were left with 370 articles.

    \item \textbf{Step 3 – Pre-processing:} We pre-processed the metadata of each article (in bib and JSON formats) and created a Comma Separated Value (CSV) file using JabRef \cite{jr}. The initial CSV file had 30 columns (fields), from which we extracted six (6) relevant fields, namely “Author”, “Title”, “Year”, “Abstract”, “Publisher”, and "NIC Type". NIC type refers to the type of SNIC used in the article, i.e.,  “DPU”, “FPGA-NIC” or “SmartNIC”. 

    \item \textbf{Step 4 – Dataset preparation:} We read each article to extract information on the type, model, and manufacturer of the SNIC(s) used by the researcher(s) / author(s). We also extracted information about the application domain (e.g., Medicine, AI, Databases, etc.), and specific focus of the research work (e.g., Energy conservation, Edge Processing, Performance evaluation and benchmarking, etc.). Finally, we included four (4) additional fields in our dataset, namely “Manufacturer”, “Device”, “Research Focus”, and “Application Domain”. In total, we had 10 fields and 370 rows in our dataset. The dataset curated from this work is available at \cite{data}.

    \item \textbf{Step 5 – Data analysis:} In this step, we performed a chronological analysis of the data using a combination of Python, Microsoft Excel and PowerBI. For each analysis, the dataset (of articles) was grouped into 3 bins, each spanning a 5-year interval - “2010–2014”, “2015–2019” and “2020–2024”. 
    
\end{itemize}

\section{Research Findings}
As stated above, the primary objective of this work was to perform a chronological analysis of the evolution of SNICs over a 15-year period, divided into 5-year bins. These bins would be used as guardrails for presenting the finds. 

In the results presented in this section, the terms “DPU”, “SmartNIC” and “FPGA-NIC” refer to the terms explicitly used by the authors of each publication. To avoid confusion between “SmartNIC” and “SNIC”, in this work, “SNIC” refers to a general term used to categorize all forms of “smart” NICs, i.e., ASIC / FPGA / SoC based NICs, as described in section 2. 

\subsection{Device Type}

\begin{figure}[]%
\begin{center}
\includegraphics[width=0.5\textwidth]{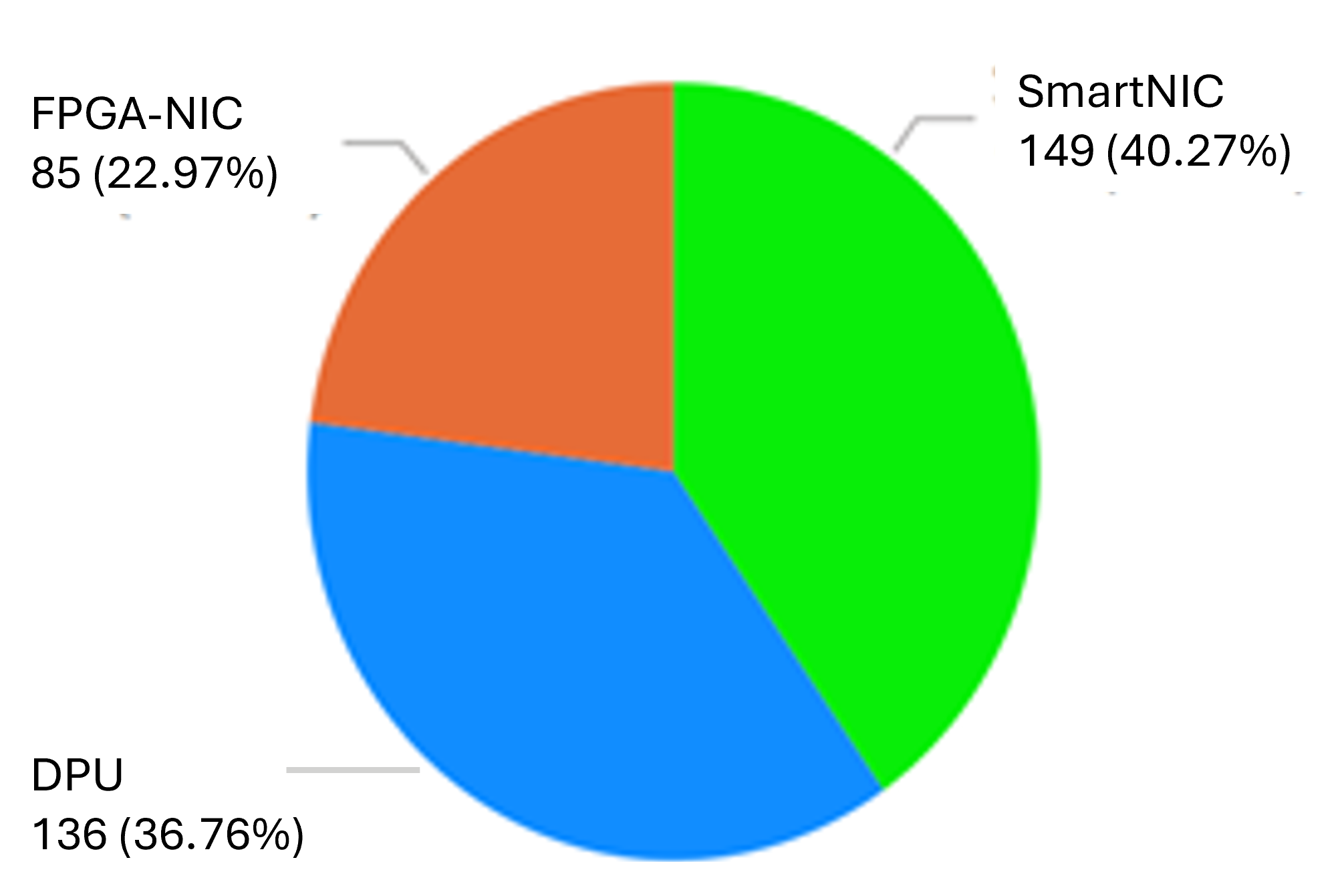}
\end{center}
\caption{Device Types \label{fig3}}
\end{figure}

\begin{figure}[h!]%
\begin{center}
\centering
\includegraphics[width=0.5\textwidth]{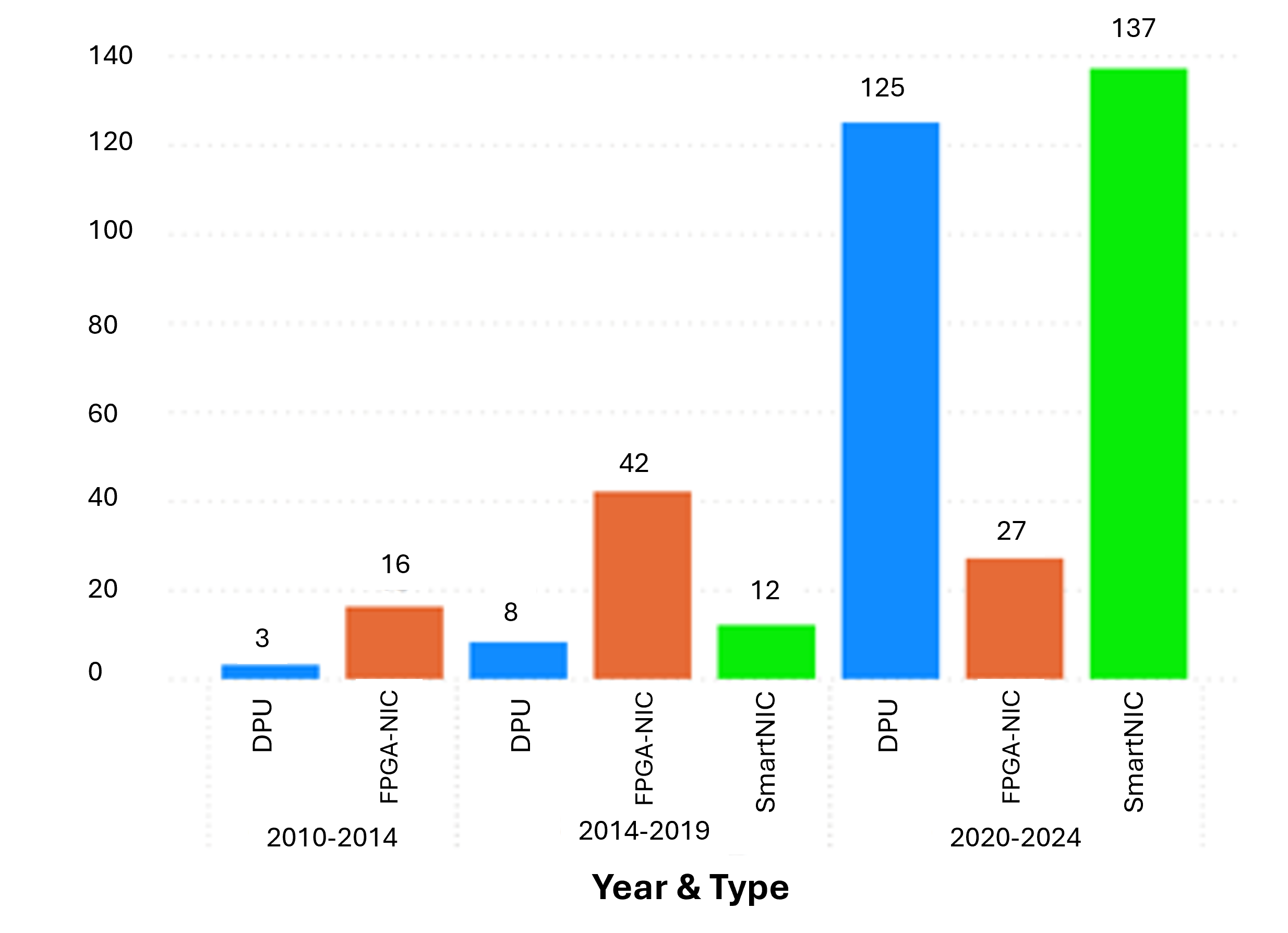}
\end{center}
\caption{Device Type distribution over the past 15 years \label{fig4}}
\end{figure}

The distribution of devices by type is shown in Fig. \ref{fig3}. Although the figure shows that “SmartNICs” are the most common, this was not the case until 2020–2024. Fig. \ref{fig4} reveals that “SmartNICs” did not exist between 2010-2014, while “DPUs” was only reported in 3 articles \cite{pmed13, bus13, rabbit14}. Also, before 2020, “FPGA-NICs” were the most common type of SNIC. The exponential rise of “DPUs” and “SmartNICs” post 2020 is likely driven by the acquisition of Mellanox by Nvidia in 2020 and the emergence of its Bluefield brand of NICs.  

\subsection{Device Manufacturers}
\begin{figure}[h!]
     \centering
     \begin{subfigure}{0.45\textwidth}
     \hspace*{.5cm}
         \includegraphics[width=\textwidth]{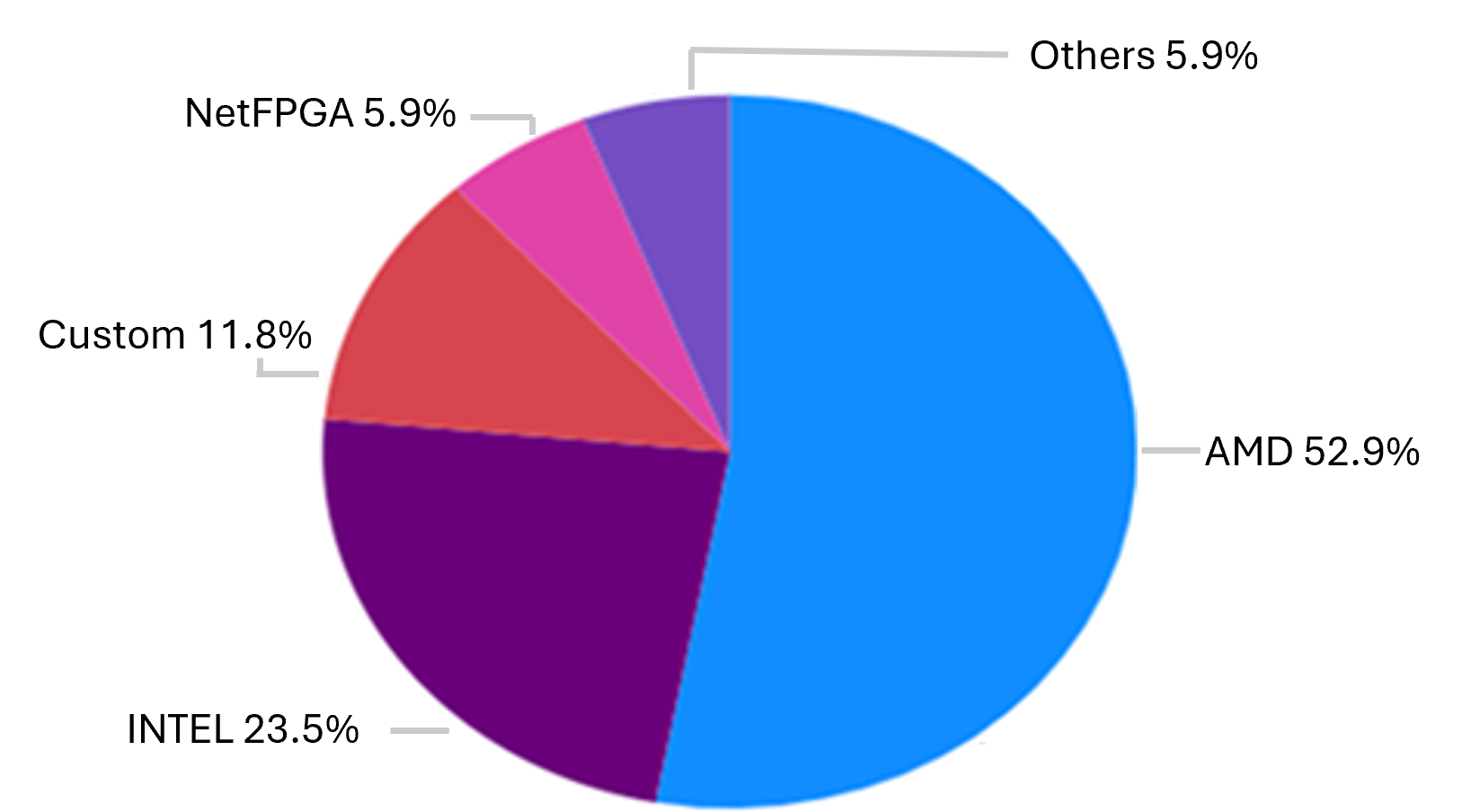}
         \subcaption{Major Manufacturers (2010–2014)}
         \label{fig5a}
     \end{subfigure}
     \begin{subfigure}{0.45\textwidth}
         \includegraphics[width=\textwidth]{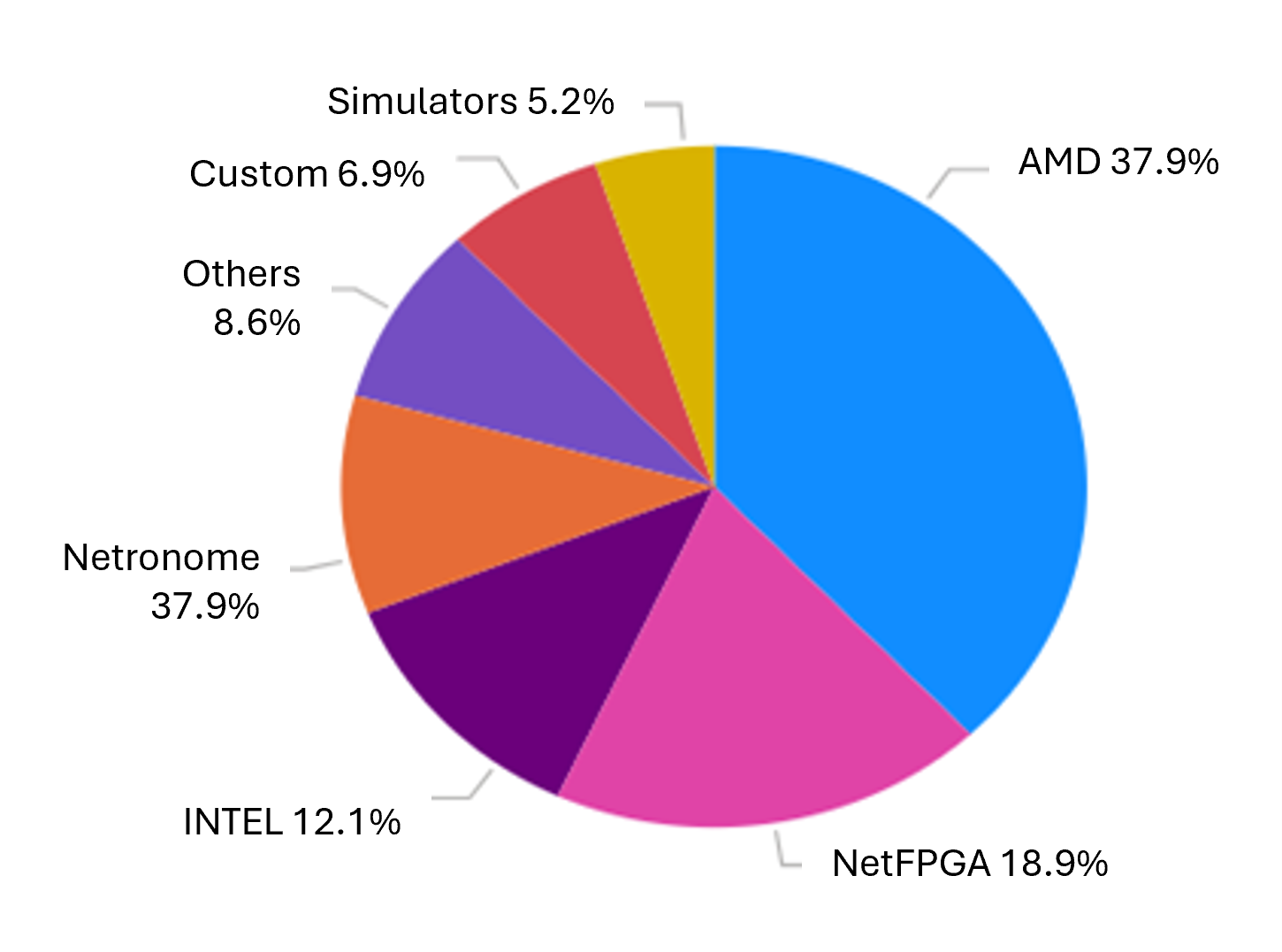}
         \subcaption{Major Manufacturers (2015–2019)}
         \label{fig5b}
     \end{subfigure}
     \begin{subfigure}{0.5\textwidth}
         \includegraphics[width=\textwidth]{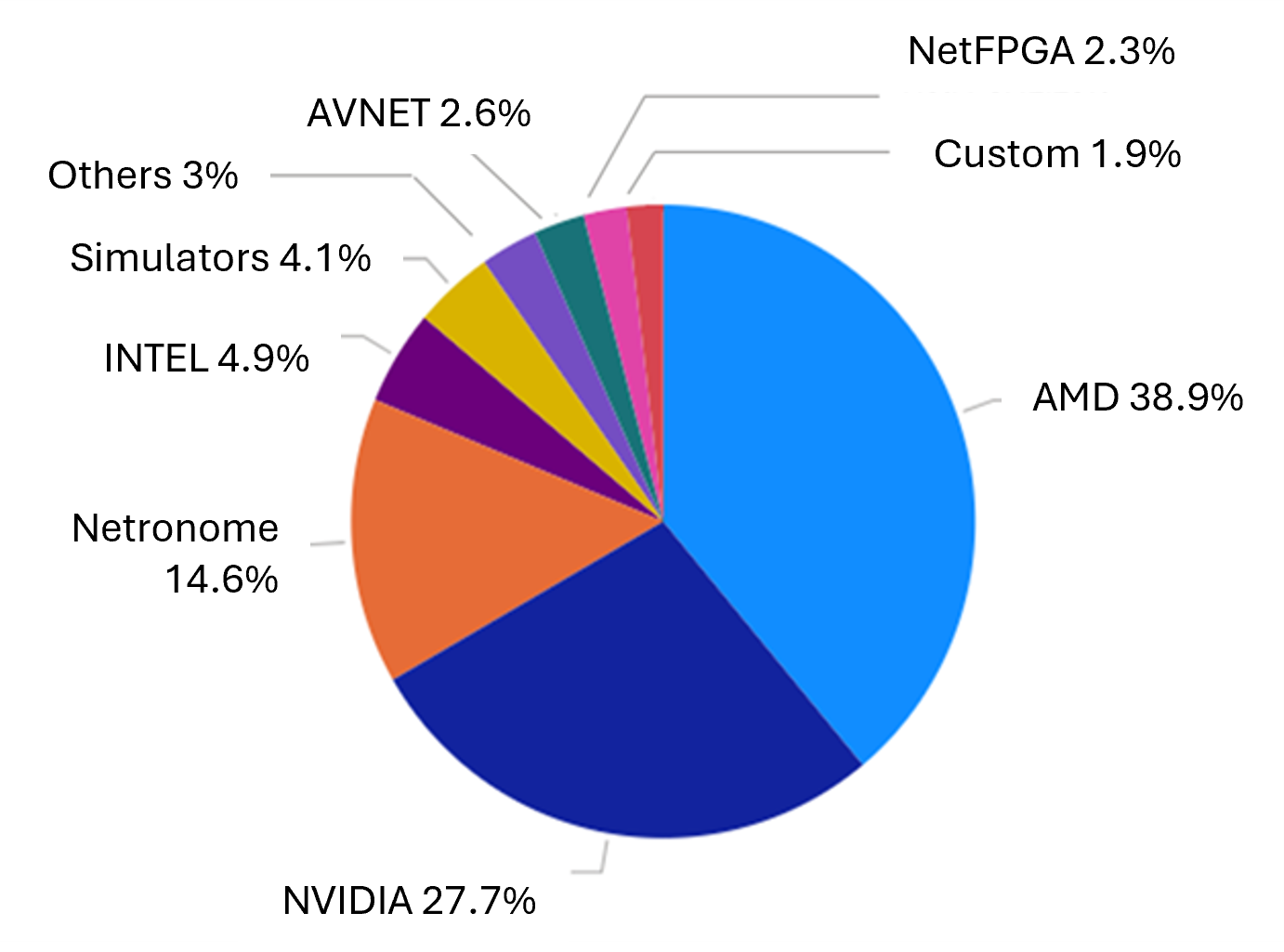}
         \subcaption{Major Manufacturers (2020–2024)}
         \label{fig5c}
     \end{subfigure}
        \caption{Device Manufacturers across the 15-year span}
        \label{fig5}
\end{figure}

During the 15-year span, Fig. \ref{fig5} shows that AMD was the dominant manufacturer. Even with the arrival of Nvidia in the 2020-2024 time frame, AMD still maintained almost 40\% of the market share. Nvidia, Netronome, and Intel were, respectively, the second, third, and fourth largest SNICs manufacturers in the past 15 years.

\subsection{Device Models}
Devices of different types and models were used in the 370 articles reviewed. In some cases, the authors provided the exact device and model used, e.g., “Netronome Agilio NFP-4000 CX Dual-Port 10 Gigabit”, while in other cases only the device family was mentioned, e.g., “Xilinx Virtex 7” or “NFP-4000”. For this reason, our report has a mix of specific device model and family.  
\begin{figure}[h!]
     \centering
     \begin{subfigure}{0.5\textwidth}
         \includegraphics[width=\textwidth]{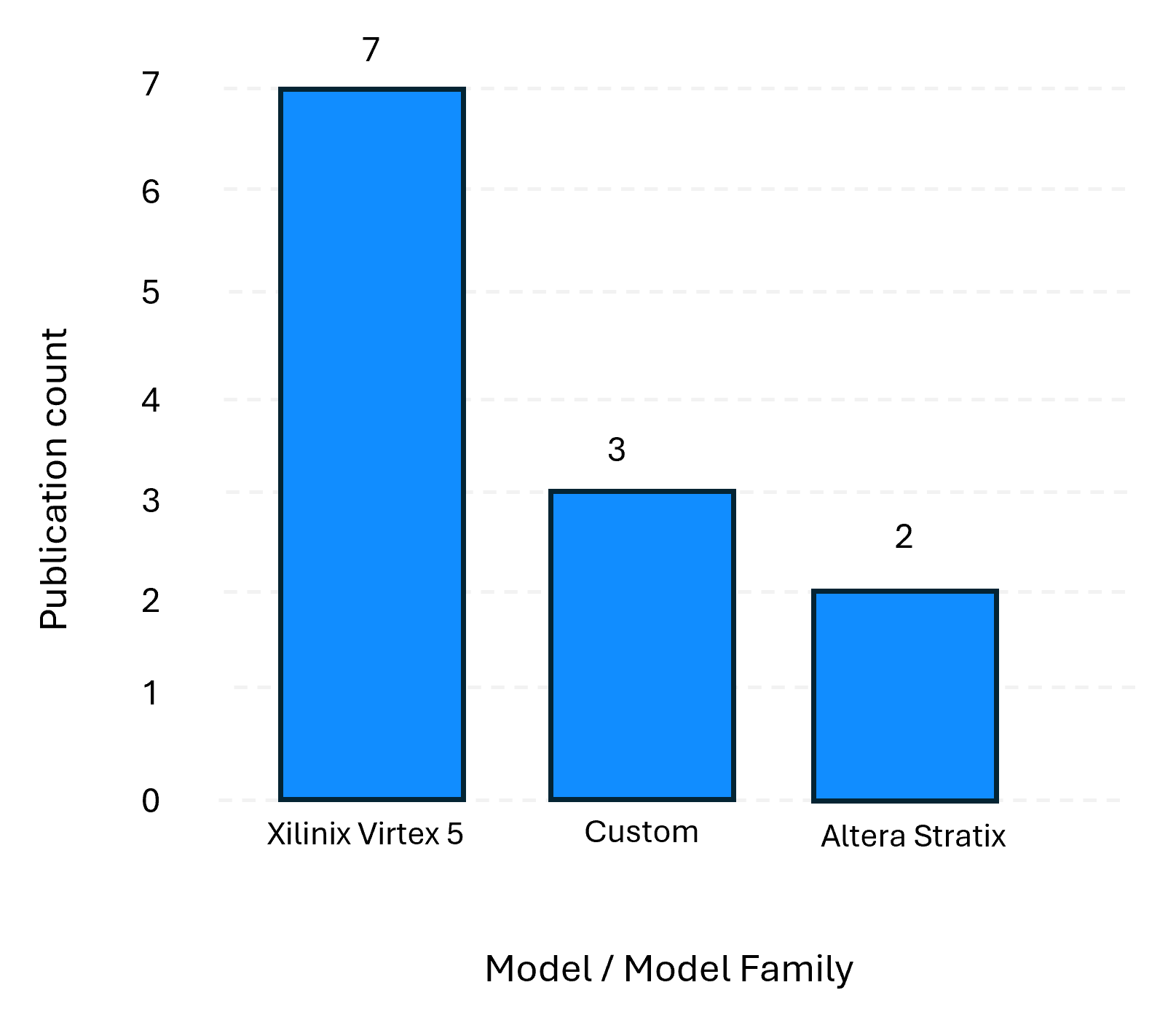}
         \subcaption{Most Common Device Models (2010–2014)}
         \label{fig6a}
     \end{subfigure}
     \begin{subfigure}{0.5\textwidth}
         \includegraphics[width=\textwidth]{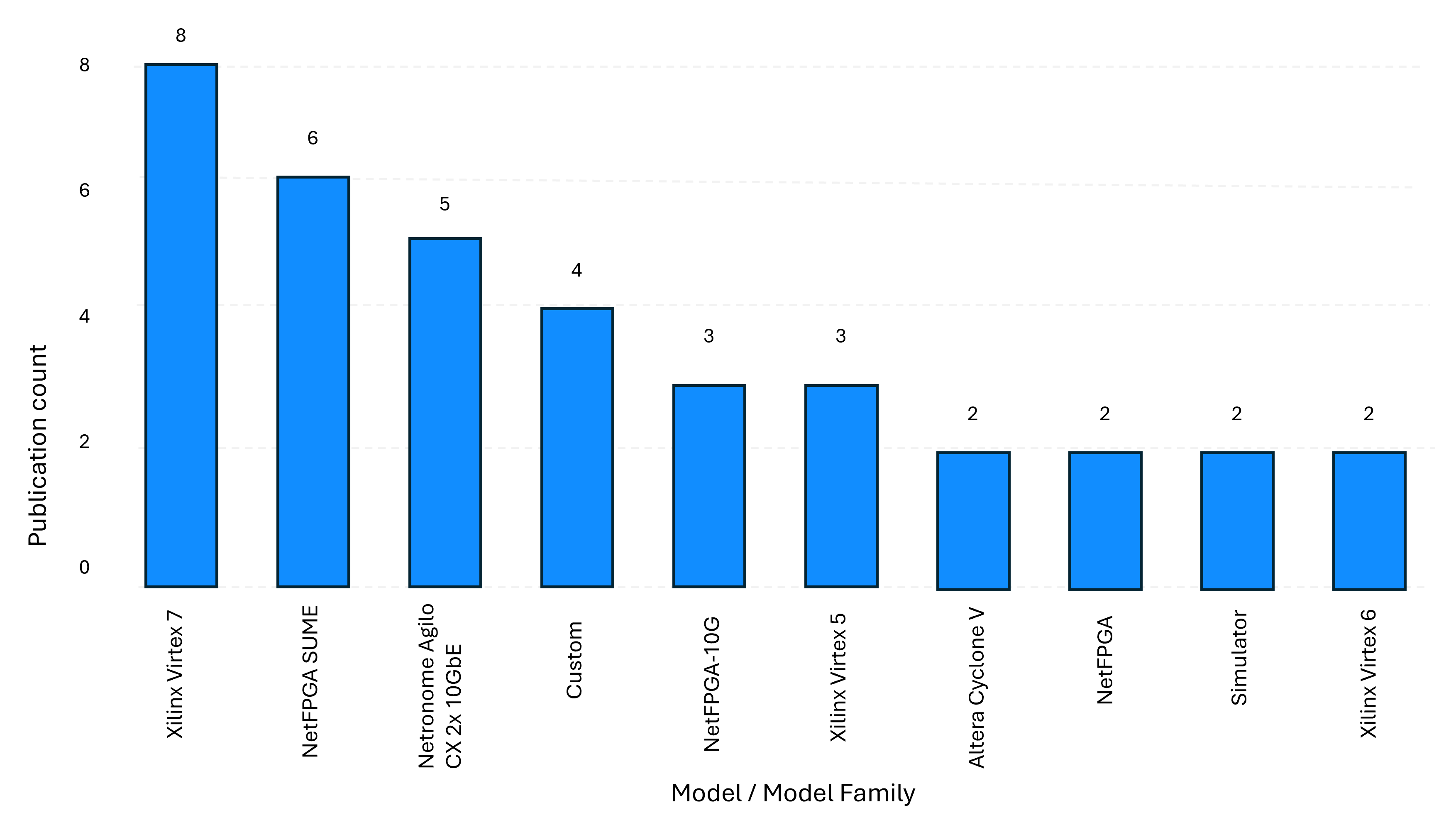}
         \subcaption{Most Common Device Models (2015–2019)}
         \label{fig6b}
     \end{subfigure}
     \begin{subfigure}{0.5\textwidth}
         \includegraphics[width=\textwidth]{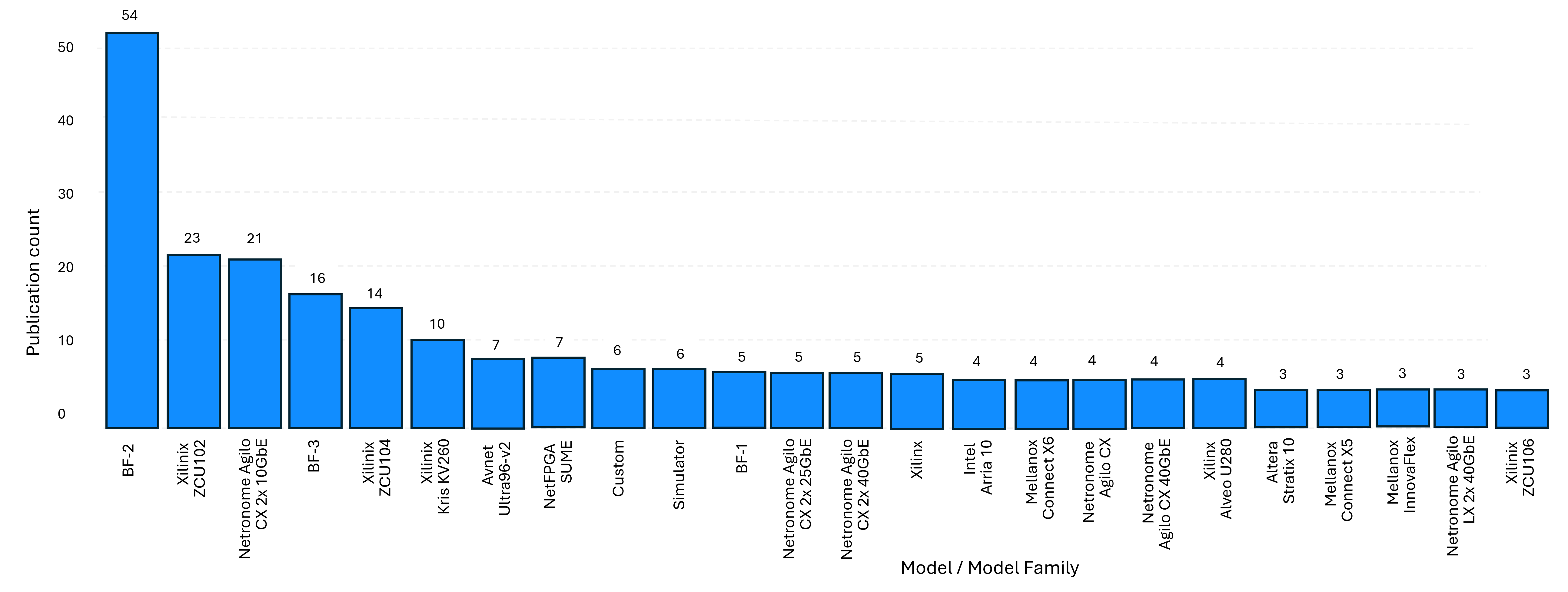}
         \subcaption{Most Common Device Models (2020–2024)}
         \label{fig6c}
     \end{subfigure}
        \caption{Top-N SNIC Devices used between 2010-2024}
        \label{fig6}
\end{figure}

Please note that the graphs shown in Fig. \ref{fig6} are truncated and only show the top-N devices for each time frame. This was done for space-saving purposes. Table \ref{tab2} provides information on the number of unique device models per time frame, using a truncation thresholds of at least 2, $(N >= 2)$. 

Between 2010 and 2014, Fig. \ref{fig6a} shows that three device models (AMD Xilinx Virtex-5, Intel Altera Stratix, and custom) were prominent. Custom refer to devices that were built by the authors, e.g., NanoBFP \cite{nano}. Other devices used in this time frame include Intel’s Altera Cyclone III, Xilinx Virtex-6, NetFPGA 10G, etc. Since each of these devices were reported in less than 2 publications, they were not shown in the graph.

Fig. \ref{fig6b} shows the top 10 devices used in the 2015-2019 time frame. AMD’s Xilinx Virtex-7 (8) was the most common, followed by NetFPGA SUME, used in 6 research works, and Netronome’s Agilio CX dual port 10GbE SmartNIC (5). Although “NetFPGA SUME” and “NetFPGA 10G” use Xilinx Virtex-7 and Virtex-5 chips, respectively, we do not consider them as part of the AMD Xilinx family. Finally, “Simulation” refers to research work in which software simulation tools, such as Vivado \cite{viva} and sPIN \cite{spin}, were used instead of physical SNICs.

The 2020-2024 time frame was dominated by “DPU”, which cumulatively represented more than 50\% of all devices. These include Nvidia’s Bluefield-2 (54) and Bluefield-3 (16), and AMD’s ZCU 102 (23) and ZCU 104 (14) DPUs. The Netronome Agilio CX 2x10GbE SmartNIC was also common, as it was used in about 21 publications. The graph also shows that 4 articles used “Netronome Agilio CX”, unfortunately, the authors did not specify the exact device model within the Agilio CX family of SmartNICs they used. Although only 24 devices are shown in Fig. \ref{fig6c}, a total of 94 unique devices were used in this period. The remaining 70 are not shown because they were used in less than 4 research work each ($N <= 3$ in Table \ref{tab2}).

\begin{table}[t]
\footnotesize
\begin{center}
\caption{Summary of device count per time frame\label{tab2}}
\begin{tabular}{|l|c|l|} \hline%
Time frame & Unique Device Models & Top N (Truncation point)\\\hline
2010 - 2014 & 11 & N: $count \leq 2$ \\\hline
2015 - 2019 & 32 & N: $count \leq 2$ \\\hline
2020 - 2024 & 94 & N: $count \leq 3$ \\\hline
\end{tabular}%
\end{center}
\end{table}

\subsection{Research Focus}
\begin{figure}
\centering
     \begin{subfigure}{0.5\textwidth}
        \includegraphics[width=\textwidth]{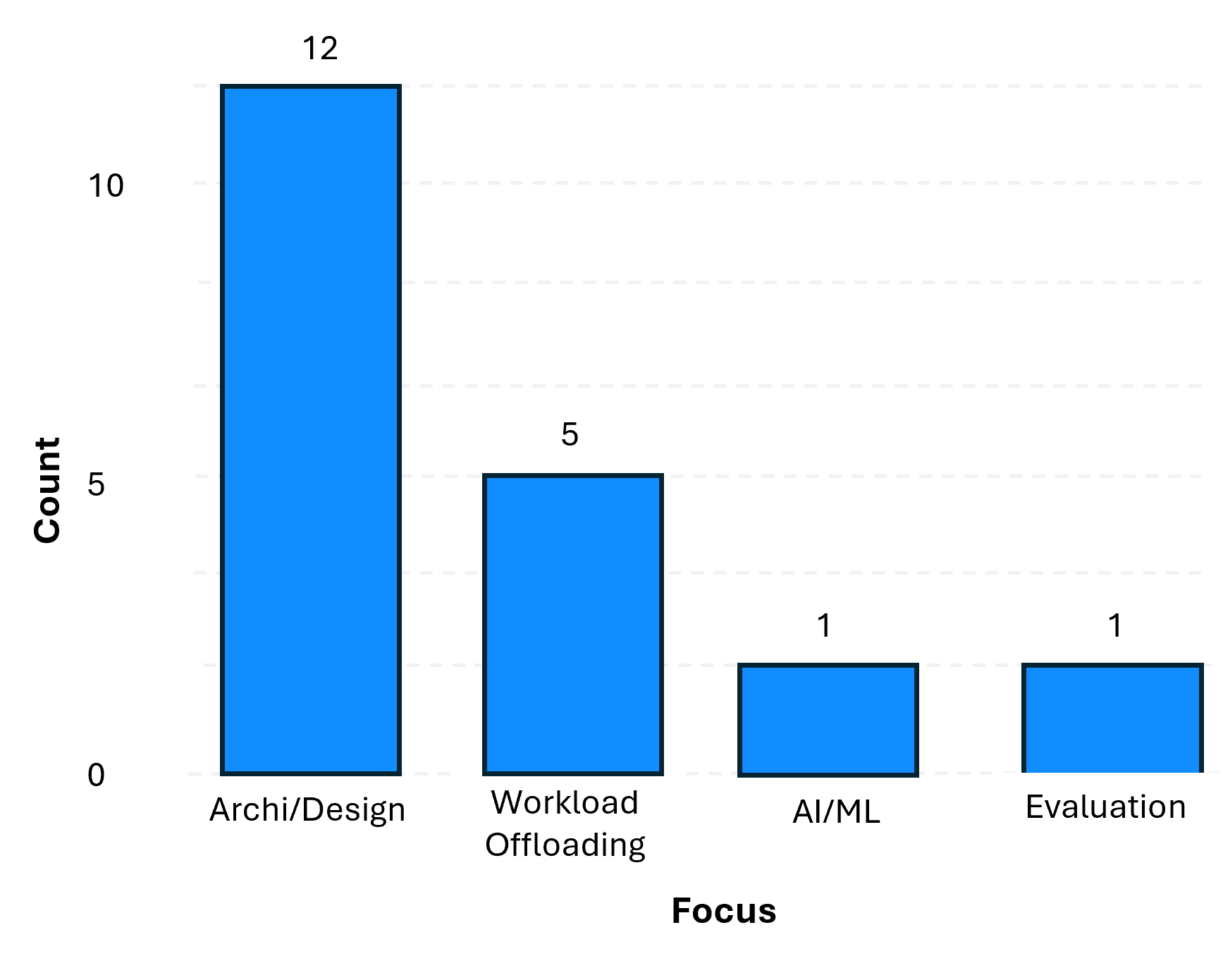}
         \subcaption{Major Research Focus (2010 – 2014)}
         \label{fig7a}
     \end{subfigure}
     \begin{subfigure}{0.5\textwidth}
         \includegraphics[width=\textwidth]{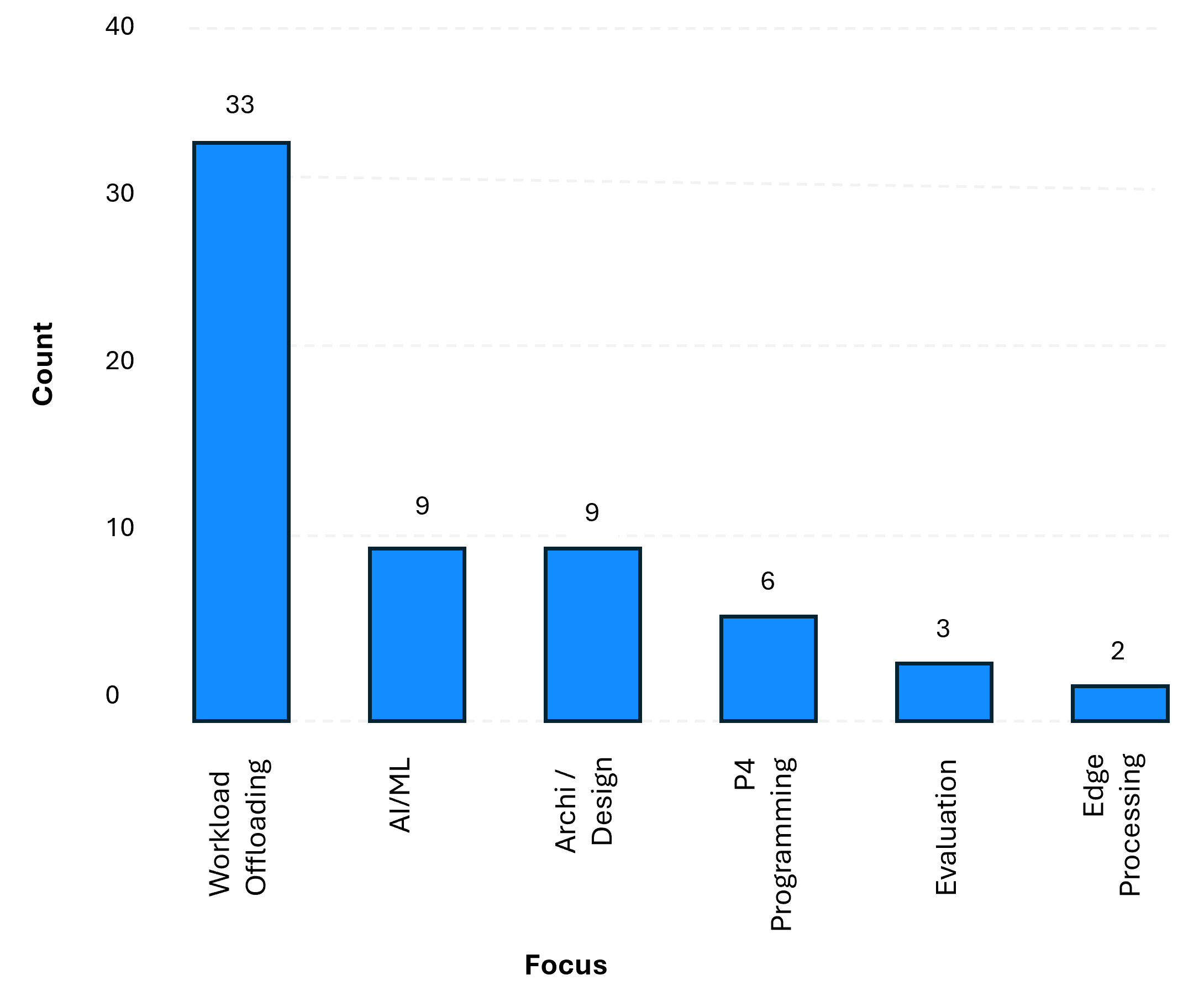}
         \subcaption{Major Research Focus (2015 – 2019)}
         \label{fig7b}
     \end{subfigure}
     \begin{subfigure}{0.5\textwidth}
         \includegraphics[width=\textwidth]{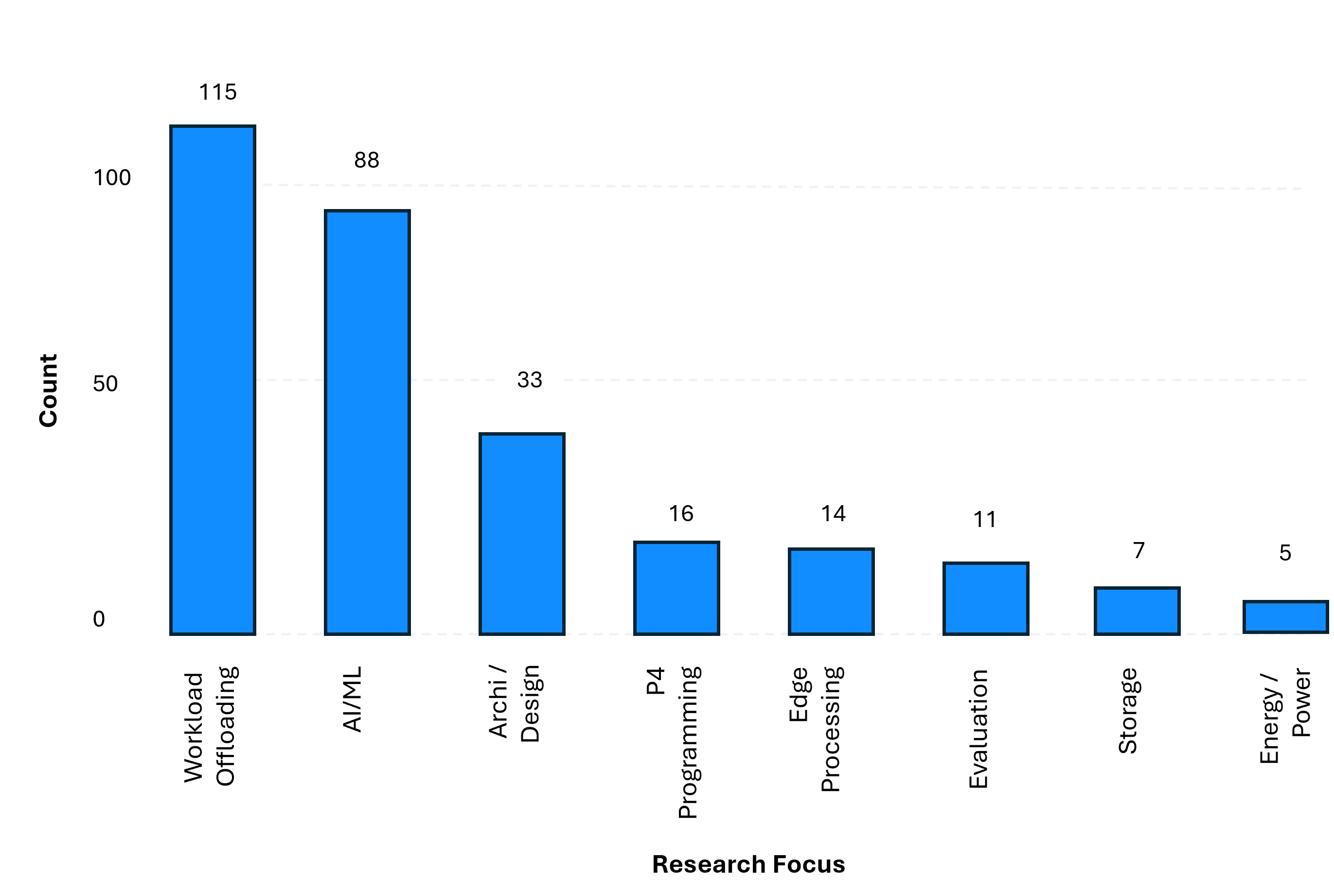}
         \subcaption{Major Research Focus (2020 – 2024)}
         \label{fig7c}
     \end{subfigure}
        \caption{Distribution of Research Focus across the 15-year span}
        \label{fig7}
\end{figure}

Fig. \ref{fig7a} shows that, between 2010 and 2014, researchers mostly focused on the architecture and design (Archi/Design) of FPGAs to improve network functionality, specifically reducing network latency. The few offloading tasks reported in this time frame were applied to Message Passing Interface (MPI) in HPC \cite{mp10}, memory (direct memory access), database processing and artificial intelligence in medical applications \cite{pmed13}.

In addition to the four research areas in 2010-2014, programming of network devices using the P4 language \cite{p4nfv, reuse19} and Edge processing \cite{mq18, uwb19} were also investigated between 2015 and 2019, as shown in Fig. \ref{fig7b}. In the same period, the offloading of workloads onto SNICs became mainstream, with most offloading tasks geared towards security \cite{chk17} and databases \cite{nosq17}. Similarly, the use of these devices to design and run Artificial Intelligence \& Machine Learning tasks (AI/ML) \cite{cnn19, dl19} also became common. 

By 2020-2024, SNICs (“SmartNIC”, “DPUs”, and “FPGA-NIC”) had become well established as tools for offloading various workloads and running AI/ML tasks. Fig. \ref{fig7c} shows that these two research areas represented 70\% (40\% and 30\% respectively) of all research work related to SNICs in that time frame. In the same period, the application of SNICs in energy and power conservation \cite{util23, hal24} as well as storage \cite{fs22, nvme24} also become popular.

\subsection{Application Domains / Use Cases}
Table \ref{tab3} provides a concise distribution of the various use cases of SNICs (“DPUs”, “FPGA-NIC”, and “SmartNIC”) between 2010 and 2024. 

\begin{table*}
\footnotesize
\centering
\caption{Applications of SNICs across various Domains from 2010 to 2024\label{tab3}}
\begin{tabular}{|p{5cm}|c|c|c|c|c|c|c|c|r|}\hline%
& \multicolumn{8}{|c|}{Years} &  \\\cline{2-10}%
& \multicolumn{2}{|c|}{2010-2014} & \multicolumn{3}{|c|}{2015-2019} & \multicolumn{3}{|c|}{2020-2024} & \\\cline{2-10}%
{Application domain / Use cases} & DPU & FPGA-NIC & DPU & FPGA-NIC & SmartNIC & DPU & FPGA-NIC & SmartNIC & Total \\\hline%

{5G / 6G / Wireless Networks} & & & & 2 & & 3 & & 10 & 15 \\\hline%
{AI/ML Specific Tasks} & & & & 3 & & 16 & 3 & 6 & 28 \\\hline%
{Autonomous Vehicles \& Automobile } & & & 1 & 2 & & 10 &  & 2 & 15 \\\hline%
{Blockchain} & & & & 2 & &  &  & 2 & 4 \\\hline%
{Cache / Memory} & & 5 & 2 & 5 & & 7 & 6 & 8 & 33 \\\hline%
{Cloud Computing} & & & &  & & 1 & 1 & 8 & 10 \\\hline%
{Data / Databases} & & 3 & 1 & 8 & & 7 & 3 & 13 & 35 \\\hline%
{Edge Processing} & & & &  & & 10 &  & 1 & 11 \\\hline%
{Energy / Power} & & & 1 & & & 3 &  &  & 4 \\\hline%
{Generic Tasks} & & & & 2 & & 2 & 2 & 3& 9 \\\hline%
{Graph Processing} & & & & & & 2 & & & 2 \\\hline%
{Human Activity Detection} & & & & & & 2 & & & 2 \\\hline%
{Medicine} & 1 & 1 & 2 & & & 6 & & 2 & 12 \\\hline%
{Molecular Dynamics} & & & &  & & 1 & & 1& 2 \\\hline%
{MPI / HPC} & & 1 & &  & & 3 & & 4& 8 \\\hline%
{Network Processes} & & 2 & & 7 & 1 & 5 & 3 & 28 & 46 \\\hline%
{Performance Benchmarking / Improvement} & 1 & 1 & & 2 & 2 & 24 & 2 & 8 & 40 \\\hline%
{Resource Management} &  &  & &  &  & 2 & 1 & 1 & 4 \\\hline%
{Robotics} &  &  & &  &  & & & 2 & 2 \\\hline%
{SDN/NFV} & & & & 2 & 5 & 1 & 1 & 11 & 20 \\\hline%
{Security} & 1 & 3 & 1& 7 & 4 & 18 & 5 & 27 & 66 \\\hline%
{Task Scheduling} & & & & & & 2 & & & 2 \\\hline%
{Total} & 3 & 16 & 8 & 42 & 12 & 125 & 27 & 137 & 370 \\
\hline
\end{tabular}%
\end{table*}

In perusing the table, it immediately becomes clear that SNICs are pervasive and have been applied in numerous domains, beyond just networking. This is due to the offloading capabilities of SNICs, specifically the “DPU” and “SmartNIC”. During the 15-year span, the highest number of non-network-specific use cases occurred in the most recent years (2020–2024).

“Security” had the highest number of applications (66 or 18\% of all articles reviewed). This is unsurprising, as these use cases were aimed at securing networks. “Network Processes” had the second highest (46) use case. Again, this comes as no surprise, as SNICs are primarily network devices, and network processes include tasks such as network compression, filtration and firewalls.  Interestingly, many researchers focused on “Performance Benchmarking / Improvement” (40), that is, analyzing and benchmarking the performance of SNICs. This implies that about 11\% of all published work focused on the design / architecture of various forms of SNICs and benchmarking these variants against the state of the art.  

Despite the popularity of AI and ML, only 7.5\% of the publications reviewed discussed the application of SNICs to AI/ML specific tasks. To avoid confusion, in this context, “AI/ML Specific Tasks” refer to tasks that are solely associated with AI/ML. These include training \cite{tn21, tn22}, quantization \cite{qnt23, qnt24}, and pruning \cite{pn20, pn24} of AI/ML models on SNICs. The limited applications of SNICs in this domain is understandable, as SNICs are resource constrained with modest storage and processing capabilities, thus, are generally ill-suited for most AI/ML specific tasks, which usually require enormous computing resources.

\section{Insight \& Discussions}
Between 2010 and 2024, the following insights can be drawn about SNICs:
\begin{itemize}
    \item FPGA-based and SoC-based SNICs were the most common type of SNICs due to their programmability and versatility.

\item FPGAs are a major technological pedestal upon which modern day SNICs are built. This is because most “DPUs” and “SmartNICs”, except Nvidia Bluefield 2 and 3, are hybridized forms of FPGA + SoC NICs. 

\item Regarding the terminology debate, the choice between “SmartNIC” versus “DPU” is mainly the manufacturer's preference. Considering the top 4 brands, Netronome uses the term "SmartNIC”, while AMD uses both terms - “SmartNIC” for its Alveo series and “DPU” for its Pensando series. Nvidia uses the term “SmartNIC” for its Connect-X series and “DPU” and/or “SuperNIC” for its Bluefield series. Intel uses the term "PAC" for its general or application-specific workload accelerators and Infrastructure Processing Unit (IPU) for its NICs developed for offloading infrastructure management tasks, such as networking, storage, and security. 

\item Many of the smaller SNIC manufacturers utilize AMD's FPGAs as their core. For example, NetFPGA’s SUME and NetFPGA-10G NICs use the Xilinx Virtex-7 and Virtex-5 FPGAs respectively. Similarly, Napatech NT200AS and NT40A01 NIC, respectively, use Xilinx XCVU5P and Xilinx KU11P FPGA, while Alpha-Delta ADM-PCIE-9H3 uses Xilinx XCVU33P. Netcope’s NFB-200G2QL uses Xilinx XCVU7P, while Hitech Global’s HTG-V5TXT is built on AMD's Xilinx Virtex-5 FPGA. 

\item The rise of “DPU” and “SmartNIC” has expanded the use of SNICs from pure networking tools to generalized devices (accelerators) capable of offloading different tasks from the host CPU. SNICs are now ubiquitous and are increasingly being applied in other seemingly unconventional application domains, such as autonomous vehicles, robotics, and medicine. 

\end{itemize}
It is important to note that the manufacturers and devices mentioned in this work do not depict the entire landscape of SNICs, as this work is based on research articles published only in IEEE Xplore repository. There are numerous other manufacturers, such as Marvell - LiquidIO SmartNICs, Broadcom - Stingray, Cisco – Nexus, etc., that were not presented in this work. In addition, we have also identified about 1,500 related publications in ACM repository, which when analyzed might buttress or skew the results presented in this work.  

\section{Conclusion}
In this paper, we performed a chronological analysis of the evolution of Smart Network Interface Cards (SNICs) over a 15-year period, from 2010 to 2024. 370 articles on SNICs published in IEEE Xplore repository were reviewed to gain insight into the evolution of SNICs and their applications in various domains. We identified the four main SNIC manufacturers, which are AMD, Nvidia, Intel, and Netronome, and their most prominent device models. We also discovered an exponential increase in the popularity of SNICs, specifically DPUs, between 2020–2024, and their general acceptance as accelerators to offload diverse workloads from the CPUs of their hosts. 

Although this work unearthed several useful insights about SNICs, it only considered articles published in IEEE Xplore. The exploration of other repositories, specifically ACM, could be a potential avenue to extend this work. Additionally, the use of SNICs in space and orbital satellites and in smart network switches was not considered in this work. These are opportunities for potential future work.

\end{document}